\documentclass{article}
\usepackage{amssymb}
\begin{document}
\title{Solutions of the Helmholtz equation given by solutions of the eikonal equation}

\author{G.F.\ Torres del Castillo \\ Departamento de F\'isica Matem\'atica, Instituto de Ciencias \\
Universidad Aut\'onoma de Puebla, 72570 Puebla, Pue., M\'exico \\[2ex]
I.\ Rubalcava Garc\'ia \\ Facultad de Ciencias F\'isico Matem\'aticas \\ Universidad Aut\'onoma de Puebla, 72570 Puebla, Pue., M\'exico}

\maketitle

\begin{abstract}
We find the form of the refractive index such that a solution, $S$, of the eikonal equation yields an exact solution, $\exp ({\rm i} k_{0} S)$, of the corresponding Helmholtz equation.
\end{abstract}

\section{Introduction}
In a recent paper \cite{Ci}, it has been shown that if a function $S$ satisfies the Laplace equation then $\exp ({\rm i} S/ \hbar)$ is an exact solution of the Schr\"odinger equation, with a velocity-independent potential $V$ (determined by $S$), if and only if $S$ is a solution of the Hamilton--Jacobi equation with the same potential $V$. As pointed out in Ref.\ \cite{Ci}, there exists a similar relation between solutions of the Helmholtz equation and solutions of the eikonal equation. The aim of this note is to give a proof of this relationship and to present some explicit examples, characterized by various refractive indices (see also Ref.\ \cite{GS}).

The Helmholtz equation is obtained from the wave equation in the case of a monochromatic wave. Substituting
\[
\Psi({\bf r}, t) = \psi({\bf r}) \, {\rm e}^{- {\rm i} \omega t},
\]
where $\omega$ is a constant, into the wave equation
\[
\nabla^{2} \Psi - \frac{n^{2}}{c^{2}} \frac{\partial^{2} \Psi}{\partial t^{2}} = 0,
\]
where $n$ is the refractive index of the medium, one finds that $\psi({\bf r})$ must obey the Helmholtz equation
\[
\nabla^{2} \psi + k^{2} \psi = 0,
\]
where
\[
k = \frac{n \omega}{c}.
\]

On the other hand, the eikonal equation,
\[
|\nabla S|^{2} = n^{2},
\]
arises in geometrical optics. The orthogonal trajectories to the level surfaces of $S$ are possible light rays in a medium with the refractive index $n$. The eikonal equation is an approximation to the Helmholtz equation, in the short wavelength limit, in the sense that if one looks for a solution of the Helmholtz equation of the form $\psi = \exp ({\rm i} k_{0} S)$ with $k_{0} \equiv \omega/c$ and
\[
S = S_{0} + \frac{1}{{\rm i} k_{0}} S_{1} + \left( \frac{1}{{\rm i} k_{0}} \right)^{2} S_{2} + \cdots ,
\]
then one finds that $S_{0}$ obeys the eikonal equation.

In Section 2 we show that if a function $S$ satisfies the Laplace equation, then $\exp ({\rm i} k_{0} S)$ is an {\em exact}\/ solution of the Helmholtz equation, corresponding to a refractive index determined by $S$, if and only if $S$ is a solution of the eikonal equation, with the same refractive index. In Section 3 we give some examples of this relationship.

\section{Sharing solutions}
We shall consider solutions of the Helmholtz equation
\begin{equation}
\nabla^{2} \psi + k^{2} \psi = 0 \label{he}
\end{equation}
of the form
\begin{equation}
\psi = \exp({\rm i} k_{0} S), \label{rel}
\end{equation}
where $S$ is a real-valued function and $k_{0}$ is a real constant. Substitution of (\ref{rel}) into Eq.\ (\ref{he}) yields
\[
\nabla^{2} \psi + k^{2} \psi = \psi \, \big( {\rm i} k_{0} \nabla^{2} S - k_{0}{}^{2} |\nabla S|^{2} + k^{2} \big).
\]
This last equation shows that, assuming that $S$ satisfies Laplace's equation
\begin{equation}
\nabla^{2} S = 0, \label{lap}
\end{equation}
$\psi$ is a solution of the Helmholtz equation (\ref{he}) if and only if $S$ is a solution of the eikonal equation
\begin{equation}
|\nabla S|^{2} = n^{2}, \label{eik}
\end{equation}
with
\begin{equation}
k = n k_{0}.
\end{equation}
({\em Cf}.\ Ref.\ \cite{Ca}.)

Instead of looking for simultaneous solutions of Eqs.\ (\ref{lap}) and (\ref{eik}) for a given refractive index (whose existence is not guaranteed), we choose a solution of the Laplace equation (containing free parameters, if possible) and {\em define} the refractive index by means of Eq.\ (\ref{eik}) ({\em cf}.\ Ref.\ \cite{Ci}).

\section{Examples}
In this section we consider some solutions of the Laplace equation and compute the corresponding refractive index. According to the result of Section 2, the proposed solutions of the Laplace equation lead immediately to solutions of the eikonal equation and of the Helmholtz equation, for that refractive index.

\subsection{Constant refractive index}
A somewhat trivial solution of the Laplace equation is given by any linear function of the Cartesian coordinates
\begin{equation}
S = a_{1} x + a_{2} y + a_{3} z, \label{linear}
\end{equation}
where $a_{1}, a_{2}, a_{3}$ are arbitrary real constants. Substituting this expression into Eq.\ (\ref{eik}) one obtains the constant refractive index equal to the norm of the vector $(a_{1}, a_{2}, a_{3})$,
\[
n = |(a_{1}, a_{2}, a_{3})|.
\]
Thus, taking into account that the refractive index depends on the norm of the vector $(a_{1}, a_{2}, a_{3})$ only, we parameterize the vector $(a_{1}, a_{2}, a_{3})$ with the aid of spherical coordinates, so that $S$ can be expressed in the form
\begin{equation}
S = n_{0} (x \sin \theta \cos \phi + y \sin \theta \sin \phi + z \cos \theta). \label{slin}
\end{equation}
In this manner, $S$ is a complete solution (because it contains two arbitrary parameters, $\theta$ and $\phi$) of the eikonal equation corresponding to the constant refractive index $n_{0}$.

Given a complete solution of the eikonal equation, for some specific refractive index, we can find all the possible light rays in a medium with that refractive index, following the same procedure as in the case of a complete solution of the Hamilton--Jacobi equation: If $S(x, y, z, P_{1}, P_{2})$ is a solution of the eikonal equation containing two independent parameters, $P_{1}, P_{2}$, then defining
\begin{equation}
Q_{i} \equiv \frac{\partial S}{\partial P_{i}} \label{qs}
\end{equation}
($i = 1, 2$), Eqs.\ (\ref{qs}) determine the light rays in terms of the four parameters $P_{1}, P_{2}, Q_{1}, Q_{2}$.

In the case of the eikonal function (\ref{linear}), identifying $P_{1}, P_{2}$ with $\theta, \phi$, Eqs.\ (\ref{qs}) give two linear equations for the coordinates $x, y, z$, each representing a plane and, therefore, their intersection is some straight line, as expected in the case of a medium with constant refractive index. The level surfaces of $S$ (the wavefronts) are planes and the corresponding solutions of the Helmholtz equation, $\psi = \exp({\rm i} k_{0} S)$, are plane waves. As is well known, any solution of the Helmholtz equation with constant refractive index can be expressed as a superposition of plane waves with constant coefficients.

\subsection{A refractive index with cylindrical symmetry}
A second example is given by the function
\begin{equation}
S = {\textstyle \frac{1}{2}} a[(x^{2} - y^{2}) \cos \theta + 2xy \sin \theta] + bz, \label{cil}
\end{equation}
where $a, b$, and $\theta$ are constants. One can readily verify that this function satisfies the Laplace equation for all values of $a, b$, and $\theta$, and that [see Eq.\ (\ref{eik})]
\begin{equation}
n = \sqrt{a^{2} (x^{2} + y^{2}) + b^{2}}. \label{ricil}
\end{equation}
Thus, the refractive index depends only on $x^{2} + y^{2}$, and it contains the constants $a$ and $b$. This means that $S$ contains one free parameter only (the angle $\theta$) and, therefore, by contrast with (\ref{linear}), is not a complete solution of the eikonal equation. Nevertheless, finding the orthogonal trajectories of the level surfaces of $S$ one obtains an infinite number of possible light rays in a medium with refractive index (\ref{ricil}).

\subsection{A refractive index with spherical symmetry}
The function
\begin{equation}
S = \frac{a}{\sqrt{x^{2} + y^{2} + z^{2}}}, \label{sph}
\end{equation}
where $a$ is a constant, is a solution of the Laplace equation, except at the origin. According to Eq.\ (\ref{eik}), the corresponding refractive index is given by
\begin{equation}
n = \frac{a}{x^{2} + y^{2} + z^{2}}. \label{risph}
\end{equation}

Since the constant $a$ appears in the refractive index, the eikonal function (\ref{sph}) does not contain free parameters and, therefore, it does not give us directly all the possible light rays in a medium with the refractive index (\ref{risph}). The wavefronts corresponding to (\ref{sph}) are spheres centered at the origin, and the orthogonal trajectories of these wavefronts are radial straight lines, which are some of the possible light rays in such a medium. The spherical symmetry of the refractive index (\ref{risph}) implies that each light ray must lie in a plane passing through the origin (in a similar way as in classical mechanics the orbits of a particle in a central field of force lie in planes passing through the center of force).

\section{Concluding remarks}
It is interesting that in the relationship between exact solutions of the Schr\"odinger equation and the Hamilton--Jacobi equation, studied in Ref.\ \cite{Ci}, and in the relationship between exact solutions of the Helmholtz equation and the eikonal equation considered here, the Laplace equation appears, but in both cases it is not clear the physical or geometrical meaning of this condition.

\section*{Acknowledgements}
One of the authors (I R-G) thanks PRODEP-SEP for financial support through a postdoctoral scholarship DSA/103.5/16/5800 and also the Sistema Nacional de Investigadores (M\'exico).


\begin{thebibliography}{9}
\bibitem{Ci} G.F.\ Torres del Castillo and C.\ Sosa S\'anchez, {\it Rev.\ Mex.\ F\'is.}\ {\bf 62} (2016) 534.
\bibitem{GS} Omar de Jes\'us Cabrera-Rosas, Ernesto Esp\'indola-Ramos, Salvador Alejandro Ju\'arez-Reyes, Israel Juli\'an-Mac\'ias, Paula Ortega-Vidals, Gilberto Silva-Ortigoza, Ram\'on Silva-Ortigoza, and Citlalli Sosa-S\'anchez, {\it J.\ Opt.}\ {\bf 19} (2017) 015603.
\bibitem{Ca} M.G.\ Calkin, {\it Lagrangian and Hamiltonian Mechanics} (World Scientific, Singapore, 1996). Chap.\ VIII.
\end{thebibliography}
\end{document}